\documentclass[%
reprint,
superscriptaddress,
amsmath,amssymb,
aps,
]{revtex4-2}

\usepackage{graphicx}
\usepackage{xcolor}
\usepackage{dcolumn}
\usepackage{bm}
\usepackage{braket}
\usepackage[none]{hyphenat}
\usepackage[colorlinks=true,linkcolor=darkblue,urlcolor=darkblue,filecolor=darkblue,citecolor=darkblue,linktocpage=true
]{hyperref}
\definecolor{darkblue}{rgb}{0.2, 0, 0.8}

\renewcommand{\[}{\begin{equation}\begin{aligned}}
		\renewcommand{\]}{\end{aligned}\end{equation}}

\begin{document}
	
	
	\title{Self Dual Black Holes as the Hydrogen Atom}
	
	\author{Alfredo Guevara}
	\email{aguevaragonzalez@fas.harvard.edu}
	\affiliation{%
		Center for the Fundamental Laws of Nature, Harvard University, Cambridge, MA 02138\\
	}
	\affiliation{
		Society of Fellows, Harvard University, Cambridge, MA 02138\\
	}
	\author{Uri Kol}%
	\email{urikol@fas.harvard.edu}
	\affiliation{%
		Center of Mathematical Sciences and Applications, Harvard University, MA 02138, USA\\
	}%
	
	\date{\today}
	
	\begin{abstract}
		Rotating black holes exhibit a remarkable set of hidden symmetries
		near their horizon. These hidden symmetries have been shown to determine phenomena such
		as absorption scattering, superradiance and more recently tidal deformations,
		also known as Love numbers. They have also led to a proposal for a
		dual thermal CFT with left and right movers recovering the entropy
		of the black hole.
		
		In this work we provide a constructive explanation of these hidden
		symmetries via analytic continuation to Klein signature. We first show that the near-horizon region of extremal black holes is a Kleinian static solution with mass $M$ and NUT charge $N$. We then analyze the self-dual solution, namely a Kerr black hole with a NUT charge $N=\pm M$. Remarkably, the self-dual solution is self-similar to its near-horizon region and hence approximate symmetries become exact: in particular, the original two isometries of
		Kerr are promoted to seven exact symmetries embedded in a conformal algebra. We analyze its full conformal group in Kleinian twistor space, where a breaking $SO(4,2) \to SL(2,\mathbb{R})\times SL(2,\mathbb{R})$ occurs due to the insertion of a preferred time direction for the black hole. 
		Finally, we show that the spectrum of the self-dual black hole is integrable and that the eigenvalue problem can be mapped exactly to the Hydrogen atom where the wavefunction is solved in terms of elementary polynomials.
		Perturbing to astrophysical
		black holes with $N=0$, we obtain a hyperfine splitting structure.

	\end{abstract}
	
	\maketitle
	
	
	\section{Introduction}
	
	Black Holes have been celebrated as ``The atom of the 21\textsuperscript{st} Century''
	since their original prediction culminated in recent remarkable observations \cite{dijkgraaf2019}. Interferometric experiments
	such as LIGO/Virgo \cite{LIGOScientific:2016lio} and EHT \cite{EventHorizonTelescope:2019dse} currently use advanced analytical modelling
	to produce and test precise predictions about astrophysical Kerr black
	holes. However, computations leading to phenomenological implications
	are usually cumbersome and can only be tracked as approximations.
	Of particular interest in this context are the Black Hole (BH) two-point function obtained via the Teukolsky equation \cite{Teukolsky:1972my,Teukolsky:1973ha,futterman_handler_matzner_1988},
	and the related geodesic equation, which have no closed solution in terms
	of known functions. They lie at the core of the observable features of black holes: they control the black hole spectrum as given
	by its quasi-normal modes (QNM), which manifest itself in the photon
	ring \cite{Yang:2012he} or in the low harmonics of a BH merger ringdown, see e.g. \cite{Berti:2009kk}.
	
	Hidden symmetries are phase space symmetries that can make a system
	solvable by introducing new constants of motion. In classical mechanics,
	the familiar Keplerian motion admits a hidden symmetry, the Laplace-Runge-Lenz
	(LRL) vector, which determines all bounded states as periodic elliptical orbits.
	The breaking of the LRL symmetry in the relativistic theory leads to the famous perihelion precession.
	In the quantum theory, the same hidden symmetry is inherited by the
	simplest orbital structure, namely the hydrogen atom. Pauli famously
	used the LRL symmetry to determine its exact spectrum \cite{Pauli:1926qpj}
	\[
	E_{n}=\frac{\hbar^{2}}{2ma_0^{2}n^{2}}
	\]
	in terms of the electron mass $m$ and Bohr's radius $a_0$.
	The symmetry manifests in the spectrum being highly degenerate, which is again corrected
	by the relativistic theory leading to hyperfine splitting.
	
	Can this structure persist in the quantum relativistic theory? A beautiful answer was provided by Caron-Huot and Henn, who used $\mathcal{N}{=}4$ SYM theory and its conformal symmetry to solve for the spectrum of massive bound states \cite{Caron-Huot:2014gia}. Now, what about
	gravitational theories? What is the analogue of the hydrogen atom for black holes? A hint
	to answer these questions emerged recently through the identification
	of a hidden conformal structure near the horizon of rotating black
	holes, given by $SL(2,\mathbb{R})\times \overline{SL(2,\mathbb{R})}$ generators \cite{Castro:2010fd}.
	As in the atom case, it can be used to constrain properties of their
	low frequency spectrum and orbiting solutions, which are conjectured
	to be dual to a thermal CFT \cite{Guica:2008mu,Bredberg:2009pv}. Moreover, a closely related conformal
	structure has been recently indicated as underpinning the absence
	of tidal deformability of Kerr, namely the vanishing of its Love numbers \cite{Charalambous:2021mea,Charalambous:2021kcz,Charalambous:2022rre} (see also \cite{Chia:2020yla,Hui:2021vcv,BenAchour:2022uqo,Hui:2022vbh}).
	The argument usually relies on the \textit{near zone approximation} $\omega(r-r_{+})\ll1$ for a perturbation $\Psi_{\omega}$, for which the wave operator in Kerr spacetime becomes the Casimir of a $SL(2,\mathbb{R})$ symmetry.
	
	
	In the following we will bridge and elucidate these developments by inspecting
	an exactly integrable instance of Kerr black holes: the self-dual point.
	
	\section{Symmetries near the Black Hole}
	
	The Kerr metric admits a one-parameter extension given by a
	NUT charge $N$. This is the gravitational analogue of magnetic charge
	in gauge theory, involving a topological Misner string \cite{Huang:2019cja}. Following
	\cite{Crawley:2021auj}, it is natural to work in $(2,2)$ Klein signature where such singularity is absent \footnote{The symmetry generators in this work can then be analytically continued back to Lorentzian signature.}. The Kerr Taub-NUT metric reads
	\[\label{eq:sdfs} 
	ds_{\textrm{KTN}}^{2}=\Sigma\left(\frac{dr^{2}}{\Delta} -d\theta^{2}\right)&-\frac{\sinh^{2}\theta}{\Sigma}(adt-\rho d\phi)^{2}  \\
	& +\frac{\Delta}{\Sigma}(dt+Ad\phi)^{2},
	\]
	where
	\[
	\Sigma& =r^{2}-(N+a\cosh\theta)^{2},\\
	\Delta  & =r^{2}-2Mr+N^{2}-a^{2}, \\
	A &= -a \, \sinh ^2 \theta -2N \, \cosh \theta,\\
	\rho &= r^2 -N^2 -a^2.
	\]
	In the Lorentzian case $\Sigma(r,\theta)=0$ determines the ring singularity while $\Delta (r)=0$ determines the horizons, located at
	\begin{equation}\label{eq:rpms}
		r_{\pm}=M\pm\sqrt{M^{2}-N^{2}+a^{2}}.
	\end{equation}
	In addition, there is a Kleinian horizon \cite{Crawley:2021auj} at $r_K{=}N{-}a\cosh \theta {+} e^{-\theta}\frac{N^2-M^2}{2a}$.
	In this signature, we can remove the Misner string singularity via the identification
	\[ \label{eq:cyc}
	t\sim t+4\pi N, \qquad \phi\sim \phi + 2\pi,
	\]
	together with the usual thermal cycle $t\sim t+ 2\pi \beta$. Indeed, the metric admits isometries along these cycles, namely
	\[\label{eq:l0s}
	L_{0}:=\partial_{\phi},\quad \bar{L}_{0}:=2N\partial_{t}.
	\]
	A motivation for our work is that the above can be extended to a
	$SL(2,\mathbb{R})\times\overline{SL(2,\mathbb{R})}$ \textit{hidden symmetry} acting
	on the near-zone wave equation for BH perturbations. Namely, for the approximated wave-equation
	$\Box_{near}\psi=0$, there exists differential operators such that $[L_{n},\Box_{near}]=[\bar{L}_{n},\Box_{near}]=0$, see
	appendix \ref{app:nearzonesym}. In the Lorentzian case this symmetry is broken by the identification $\phi \sim \phi + 2\pi$, under which 
	\[\label{eq:mdn}
	L_{n} & \to e^{4\pi^{2}inT_{R}}L_{n},  \\
	\bar{L}_{n} & \to e^{4\pi^{2}inT_{L}}\bar{L}_{n}, 
	\]
	where $n=-1,0,+1$ and
	\[\label{eq:temps}
	T_{R}=\frac{r_{+}-r_{-}}{4\pi a}\,,\quad T_{L}=\frac{M^{2}-N^{2}}{2\pi aM}
	\]
	are conjectured to be the temperatures of a dual thermal 2d state \cite{Castro:2010fd,Castro:2013kea, Perry:2022udk,Compere:2012jk}. In practice,
	this means that global solutions do not form conformal multiplets
	since the identification leads to singularities. A particular modification
	of $\bar{L}_{\pm1}$, dubbed Love symmetry \cite{Charalambous:2021kcz,Charalambous:2021mea,Charalambous:2022rre}, has been proposed as a
	globally defined symmetry group which leads to well defined solutions. The corresponding wave
	operator is given by its Casimir
	\begin{align}
		\Box_{\textrm{Love}} & :=\frac{\bar{L}_{+}'\bar{L}_{-}'+\bar{L}_{+}'\bar{L}_{-}'}{2}-\bar{L}_{0}'^{2}-\textbf{J}^2\nonumber \\
		& =\partial_{r}\Delta\partial_{r}-\frac{\rho^{2}(r_+)}{\Delta}\left[ \vphantom{\int_1^2} 
		(\partial_{t}+\Omega_{+}\partial_{\phi})^{2} \right. \nonumber \\
		& \left. +4\Omega_{+}\frac{r-r_{+}}{r_{+}-r_{-}}\partial_{t}\partial_{\phi}\right]-\textbf{J}^2,\label{eq:lveq}
	\end{align}
	where $\Omega_{+}=a/\rho(r_+)$ is the angular velocity at $r_{+}$. Here $\textbf{J}^2$
	is an angular operator with spheroidal eigenfunctions. The authors of \cite{Charalambous:2021kcz,Charalambous:2021mea,Charalambous:2022rre}
	used a particular near-zone approximation to show that the wave equation reduces
	to $\Box_{\textrm{Love}}\Psi=0$. The solutions assemble into polynomial representations
	of $\overline{SL(2,\mathbb{R})}$. Quite remarkably, this implies
	the vanishing of the static tidal response under perturbations of the black hole, i.e. Love numbers. We will come back to this and explain this symmetry in the next section.
	
	The situation $T_{R}\to0$ has been considered extensively as it corresponds
	to near extremal black holes, see e.g. \cite{Guica:2008mu,Bredberg:2009pv,Castro:2013kea,Compere:2012jk}. In this case the conformal symmetry is unbroken and one finds a $SL(2,\mathbb{R})$
	\textit{isometry} $\{L_{n}\}$, which acts globally near the horizon
	of Kerr. What is the origin of this isometry? We show in Appendix \ref{sec:nenhks} that as we take $T_R\to 0$ and approach $r\to r_+$ in the KTN metric \eqref{eq:sdfs} we get the following near-extremal near-horizon metric:
	\begin{equation}\label{ec:nnsk}
		ds_{\textrm{NNHEK}}^{2}=(\tilde{r}^2-M^2)\left(\frac{d\tilde{r}^{2}}{\tilde{\Delta}}-\underbrace{\left(\sigma_{+}\sigma_{-}+f(\tilde{r})\sigma_{0}^{2}\right)}_{\textrm{wAdS}_3}\right)
	\end{equation}
	where $\sigma_{m}$ are one-forms ($m=-1,0,1$) and
	\begin{equation}
		\tilde{\Delta} = \tilde{r}^2 -2\tilde{r}N + M^2\,,\quad f=\frac{4M^2 \tilde{\Delta}}{(\tilde{r}^2-M^2)^2}\,.\end{equation}
	This is nothing but the static ($a=0$) Taub-NUT spacetime with $M$ and $N$ interchanged! (See Appendix \ref{sec:nenhks}). For instance, for pure Kerr ($N=0$) the role of the NUT charge is played by the maximal spin $a\to M$. Furthermore, we can now understand the symmetry: The term in brackets is known as \textit{warped} $AdS_{3}$ space with warp factor $f$, where usual $AdS_3$ is obtained when $f=1$. Hence $\{L_{n}\}$ emerges simply because the left-isometries of $AdS_{3}$ preserve the one-forms, $[L_n,\sigma_m]=0$, and hence are isometries of \eqref{ec:nnsk} irrespective of the warp. 
	
	The above observation provides a strong motivation to understand the implications of NUT charge. In this work we observe that the analytic continuation to Klein signature allows us to explore
	a new possibility with unbroken symmetries, namely $T_{L}\to0$, corresponding to $N\to \pm M$. This is the self-dual (SD)
	Kerr Taub-NUT metric recently studied in \cite{Crawley:2021auj}. Remarkably, in this
	space the spin parameter is irrelevant and given by a coordinate transformation; thus the self-dual point can be regarded as a special case of extremal BH. Indeed, using \eqref{eq:rpms} and \eqref{eq:temps} one finds $T_R=\frac{1}{2\pi}$ which is equivalent to $T_R=0$, see  \eqref{eq:mdn} \footnote{We can get $T_R=0$ by first taking the extremal limit $r_+\to r_{-}$, followed by the SD limit $M\to N$ }.
	As we will see, this means that this spacetime enjoys an emergent rotational invariance.

	Furthermore, the generators $\{\bar{L}_{n}\}$ in (\ref{eq:mdn})
	are also now globally defined at $T_L=0$. In fact, in appendix \ref{app:nearzonesym} we check that they coincide with the Love symmetry
	$\bar{L}_{n}\approx\bar{L}_{n}'$ up to terms irrelevant at low energies. It turns out that $\overline{SL(2,\mathbb{R})}$ arises as an \textit{exact} symmetry of the self-dual wave operator, rather than an isometry, as we show.

	\section{The Self-Dual Black Hole}
	
	The fact that $T_{L}=T_{R}=0$ suggests that the SD point is the ground state of the holographic CFT  discussed in \cite{Guica:2008mu,Compere:2012jk}. Indeed, we will see that the Hamiltonian $H^{SD}$ on this background corresponds to the four-dimensional harmonic oscillator, which is invariant under a $SL(2,\mathbb{R})\times SL(2,\mathbb{R})$ group. The vacuum state of this oscillator are nothing but the static ``Love symmetry'' solutions. 
	In the following we will provide an explicit construction of these symmetries and its implications.
	
	Recent developments in scattering amplitudes have shed new light on properties of rotating black holes \cite{Guevara:2017csg,Guevara:2018wpp,Guevara:2020xjx,Guevara:2021yud,Crawley:2021auj,Chung:2018kqs,Maybee:2019jus,Kim:2023aff,Kim:2023vgb,Arkani-Hamed:2019ymq,Aoude:2021oqj}, resonating with Penrose's seminal treatment of spacetimes in terms of spinors \cite{Penrose:1985bww} and twistors \cite{penrosenlg}. The latter can indeed be used to construct a large class of SD metrics. Motivated by these ideas, here we implement a spinor-like coordinate system at the SD point. It is given by
	\[\label{eq:spinors}
	z_{+} & =\sqrt{r-M+a}\cosh\frac{\theta}{2}\,e^{\frac{it}{4M}}e^{-\frac{i\phi}{2}},\\
	z_{-} & =\sqrt{r-M-a}\sinh\frac{\theta}{2}\,e^{\frac{it}{4M}}e^{+\frac{i\phi}{2}},
	\]
	together with their complex conjugates $\bar{z}_{\pm}$. Here $z=(z_+,z_-)$ transforms as a spinor and $\pm$ denotes the weight under $L_0=\partial_\phi$. The thermal cycle becomes $t\sim t + 4\pi M$, i.e. $\beta=2M$. Indeed, thanks to the identifications \eqref{eq:cyc} these coordinates are a double cover of spacetime.
	
	Let us denote $z\bar{z}\equiv z_{+}\bar{z}_{+}-z_{-}\bar{z}_{-}=r-M+a\cosh\theta$,
	which vanishes at the Kleinian horizon $r=r_K$. Let us now present one of our main results. We define the following basis of one-forms
	\[
	\sigma_0 &= \frac{z_+ \bar{z}_+ d\log\frac{z_+}{\bar{z}_+}-z_- \bar{z}_- d\log\frac{z_-}{\bar{z}_-}}{z \bar{z}},\\
	\sigma_+ & =\bar{\sigma}_-= \frac{2z_+ z_-}{z\bar{z}} d\log \left(\frac{z_-}{z_+}\right) \label{eq:ofor}
	\]
	Now consider the metric \eqref{ec:nnsk} at the SD point, with warp $f=\frac{4M^2}{(\tilde{r}+M)^2}$ - using \eqref{eq:ofor} in \eqref{ec:nnsk} we recover precisely the KTN metric \eqref{eq:sdfs}. That is
	\begin{equation}
		ds^2_{\textrm{KTN}}= ds^2_{\textrm{NNHEK}} \,,\quad \Delta = \tilde{\Delta}\,,\quad \textrm{at}\quad N=M\,. \label{eq:sdp}
	\end{equation}
	together with the identification $\tilde{r}=r+a\cosh\theta=z\bar{z}+M$. In other words, we find that the self-dual black hole is \textit{self-similar} as we zoom into its horizon. This implies that the near horizon symmetries are exact rotational isometries. As we elaborate in appendix \ref{sec:nenhks}, this self-similarity diffeomorphism is a composition of the near-horizon diffeomorphism \cite{Bardeen:1999px} and the boost diffeomorphism of \cite{Crawley:2021auj}.  Indeed, note that the spin parameter $a$ has been absorbed
	by the coordinates $z_\pm, \bar{z}_\pm$!
	
	One can check from \eqref{eq:ofor} that the forms $\{\sigma_n\}$ are neutral under $L_0=\partial_\phi$ but charged under $\bar{L}_0=2M\partial_t$. More generally, let us introduce the $SL(2,\mathbb{R})\times\overline{SL(2,\mathbb{R})}$ isometries of $AdS_3$ which satisfy
	\begin{equation}
		[L_{n},\sigma_{m}]=0,\quad[\bar{L}_{n},\sigma_{m}]=(n-m)\sigma_{n+m}.
	\end{equation}
	It follows that $\{L_{n}\}$ and $\bar{L}_{0}=\partial_{t}$
	are four isometries of \eqref{eq:sdp}, reflecting the fact that $L_{0}=\partial_{\phi}$ can be enlarged
	to a full rotational symmetry at the SD point for any
	spin. Now, to recover the six symmetries as isometries of \eqref{ec:nnsk} we would need the warp factor $f=\frac{4M^2}{(\tilde{r}+M)^2}\to 1$. This is what happens at the Kleinian horizon $r\to M - a\cosh\theta$, namely $z\bar{z} \to 0$. Indeed, in this region the space
	becomes Klein space 
	\begin{equation}
		ds^{2}\to8N(dz_{+}d\bar{z}_{+}-dz_{-}d\bar{z}_{-}),\label{eq:klein}
	\end{equation}
	which restores the $SL(2,\mathbb{R})\times\overline{SL(2,\mathbb{R})}$
	isometry. The second result of our paper is that, under the transformation
	(\ref{eq:spinors}) the Klein Laplacian remarkably becomes 
	\begin{equation}
		\partial_{+}\bar{\partial}_{+}-\partial_{-}\bar{\partial}_{-}=\frac{\Box_{\textrm{Love}}}{r-M+a\cosh\theta}\label{eq:svs}
	\end{equation}
	as given in (\ref{eq:lveq}) for the SD point. Hence the near-zone wave equation
	here occurs near the Kleinian horizon (rather
	than the Lorentzian one). The Klein wave equation has been recently
	addressed in \cite{Atanasov:2021oyu}, in terms of the $SL(2,\mathbb{R})\times\overline{SL(2,\mathbb{R})}$ Casimirs (the $AdS_3$ generators are given there explicitly). Since $\Box_{\textrm{Love}}$ is also given by Casimirs \eqref{eq:lveq}, \eqref{eq:svs} shows that the Love generators $\{\bar{L}_n' \}$ are related to our $AdS_3$ generators by a conformal transformation. Indeed $r{-}M{+}a\cosh\theta=\tilde{r}{-}M$ is precisely the conformal factor appearing in the metric \eqref{ec:nnsk}. Since this argument is independent of the particular warp factor, we expect that it extends away from the SD point, which we will address in future work.
	
	Highest weight (static) solutions of \eqref{eq:svs} were found in \cite{Atanasov:2021oyu} to be homogeneous functions of the radius $\rho=z\bar{z}$.
	This is consistent with the vanishing of static Love numbers, namely the absence of $1/\rho$ singularities indicates no tidal response for black holes \cite{Charalambous:2021kcz}.
	
	Quite interestingly, we find that this Love symmetry persists
	exactly as a symmetry of the exact wave equation at finite $\omega$,
	albeit $\overline{SL(2,\mathbb{R})}$ is now hidden. In fact, consider
	the eigenspace
	\begin{equation}
		\bar{L}_{0}=2M\omega\equiv\mu\label{eq:con}
	\end{equation}
	The exact wave equation in the coordinates \eqref{eq:spinors} for the SD black hole is
	\begin{equation}
		H^{SD}=\partial_{+}\bar{\partial}_{+}-\partial_{-}\bar{\partial}_{-}+\omega^{2}(z_{+}\bar{z}_{+}-z_{-}\bar{z}_{-})=4M\omega^{2}.\label{eq:con2}
	\end{equation}
	This is nothing but the rectangular 4D harmonic oscillator with potential
	$\omega^{2}z\bar{z}$. The vacuum eigenspace $\bar{L}_{0}=0$ of this
	problem are the Love symmetry solutions of (\ref{eq:svs}). But even for finite
	$\omega$, it is easy to check that 
	\begin{equation}
		[L_{n},H^{SD}]=[\bar{L}_{n},H^{SD}]=0.
	\end{equation}
	Is this system integrable at finite $\omega$? Famously, the 4d oscillator can be mapped
	to the hydrogen atom, whose dynamics is controlled by a larger $SO(4,2)$ hidden conformal algebra. Using this connection, we will now explicitly realize the LRL vector in full GR as anticipated.
	
	\subsection{Hidden Symmetries from Twistors}
	
	One is tempted to employ the Love symmetry to classify the dynamical spectrum
	of $H^{SD}$. However, the action of $\bar{L}_{\pm}$ violates the
	constraint (\ref{eq:con}) so it does not take solutions to solutions. Luckily, the harmonic oscillator has a
	more natural, exact, symmetry that also commutes with $\bar{L}_{0}$.  
	Motivated by twistor description of SD spacetimes, let us examine the larger symmetry structure by promoting the spinor coordinates to twistors \footnote{This analysis is tied to position space so it is rather different than the approach of e.g. \cite{penrosenlg}.}. Consider the phase space transformation $z_{\pm}\to\frac{z_{\pm}}{\sqrt{\omega}}$.
	In these reescaled coordinates we introduce the twistor variable
	\begin{equation}
		Z=(z_{+},z_{-},\frac{\partial}{\partial\bar{z}_{-}},-\frac{\partial}{\partial\bar{z}_{+}})
	\end{equation}
	From \eqref{eq:spinors} we see that it diagonalizes $\bar{L}_0$ but not $L_0$. Indeed, it is a straightforward to check that the equations (\ref{eq:con})-(\ref{eq:con2})
	become
	\[
	\mu=\bar{L}_{0} & =Z^{A}\frac{\partial}{\partial Z^{A}}
	\\& =z_{+}\partial_{+}-z_{-}\partial_{-}+\bar{z}_{+}\bar{\partial}_{+}-\bar{z}_{-}\bar{\partial}_{-}\,,\\
	\mu=H & =Z^{A}(\gamma_{0})_{\,A}^{B}\frac{\partial}{\partial Z^{B}}
	\\ & =\partial_{+}\bar{\partial}_{+}-\partial_{-}\bar{\partial}_{-}+(z_{+}\bar{z}_{+}-z_{-}\bar{z}_{-}).
	\]
	Thus, $\bar{L}_{0}$ and $H$ are simply linear transformations in $Z^{A}$
	space! We can associate $\bar{L}_{0}=\mathbb{I}$ and $H=\gamma_{0}$.
	The question of hidden symmetries is now reduced to finding a linear
	transformation in this 4D space that commutes with $\gamma_{0}$ and
	$\mathbb{I}$. This is easy as all $4\times 4$ linear maps are spanned by Gamma
	matrices. One quickly finds
	\begin{equation}
		K_{i}=\gamma_{i}\gamma^{*}\,,
	\end{equation}
	together with $L_{i}=\gamma_{0}K_{i}$ \footnote{We warn that commutators can be interpreted as operations in twistor
		space but products such as $\gamma_{0}K_{i}$ cannot. However, this
		representation is useful to derive the algebra.}. Remarkably, $K_{i}$ is indeed the LRL vector and $L_{i}$ is the
	angular momentum of the problem: For instance, they satisfy
	
	\begin{equation}
		[K_{i},K_{j}]=[L_{i},L_{j}]=\gamma_{ij}=\epsilon_{ijk}\gamma_{k}\gamma_{0}\gamma^{*}=\epsilon_{ijk}L_{k}
	\end{equation}
	
	It follows that
	\begin{equation}
		A_{i}^{\pm}=\frac{L_{i}\pm K_{i}}{2}=\left(\frac{1\pm\gamma_{0}}{2}\right)K_{i}
	\end{equation}
	form a $SL(2,\mathbb{R})\times SL(2,\mathbb{R})$ hidden symmetry
	algebra that preserves $L_{0}$ and $H$. However, $A_i^\pm$ are only linear
	in twistor space and thus a priori differs from the Love symmetry. 
	
	All commutation relations follow from the Clifford algebra, which
	spans a $SU(2,2)\sim SO(4,2)$ full algebra \cite{Mack:1969dg}. In preparation for the
	next section, let us also introduce $X_{\mu}=\gamma_{\mu}\gamma_{+}$
	and $P_{i}=\gamma_{i}\gamma_{0}$ (recall $\gamma_{\pm}=\frac{1\pm\gamma^{*}}{2}$
	is a chiral projector) which satisfy canonical commutations
	\begin{equation}
		[X_\mu,X_\nu]=0,[X_{i},P_{j}]=\delta_{ij}X_{0}, [P_i,P_j] = \epsilon_{ijk}L_k
	\end{equation}
	It is also straightforward to check that $K_{i}=\epsilon_{ijk}P_{j}L_{k}+X_{i}$
	which is sometimes used as a definition of the LRL vector.
	
	For clarity we unpack here some operators explicitly. Undoing the
	scaling $z_{\pm}\to\sqrt{\omega}z_{\pm}$ we find, up to normalization,
	\[
	K_{i} & =Z\gamma_{i}\gamma^{*}\frac{\partial}{\partial Z}\to-\sigma_{i}^{\alpha\dot{\beta}}\partial_{\alpha}\bar{\partial}_{\dot{\beta}}+\omega^{2}\sigma_{i}^{\alpha\dot{\beta}}z_{\alpha}\bar{z}_{\dot{\beta}}, \\
	L_{i} & =Z\gamma_{0}\gamma_{i}\gamma^{*}\frac{\partial}{\partial Z}\to\sigma_{i0}^{\alpha\beta}z_{\alpha}\partial_{\beta}+\sigma_{i0}^{\dot{\alpha}\dot{\beta}}\bar{z}_{\dot{\alpha}}\bar{\partial}_{\dot{\beta}} ,\\
	X_{i} & =Z\gamma_{i}\gamma_{+}\frac{\partial}{\partial Z}\to\sigma_{i}^{\alpha\dot{\beta}}z_{\alpha}\bar{z}_{\dot{\beta}}=x_{i}.
	\]
	Here $L_{i}$ turns into the isometries of (\ref{eq:sdp}), the $x_{i}$
	turn into rectangular coordinates for (\ref{eq:sdfs}) and the $K_{i}$
	turn into a quadratic differential operator. On the subspace $\bar{L}_{0}=2N\omega$,
	we expect that $K_{i}$ coincide with the triplet of Killing tensors
	$\nabla_{\mu}K_{i}^{\mu\nu}\nabla_{\nu}$ that have appeared previously
	in the literature, but we have not managed to check this explicitly.

	\section{Hydrogen Atom and Hyperfine Splitting}
	
	In order to make an explicit connection with the hydrogen atom, let us now consider the spherical form of the wave equation on \eqref{eq:sdfs}. We will set the spin parameter to zero $a=0$, keeping in mind that it is simply a gauge choice. The wave operator takes the following separable form
	\[
	\Box & = \frac{1}{r^2-N^2}\partial_r\Delta\partial_r
	\\ &
	-\left(4N^2 - \frac{(r^2-N^2)^2}{\Delta} \right)\partial_t^2
	+\textbf{J}^2,
	\]
	where $\textbf{J}^2$ is the angular momentum operator (whose explicit form is given in \eqref{angularSquare}). The eigenfunctions of $\textbf{J}^2$ are the \emph{monopole harmonics} (also known as the \emph{generalized spherical harmonics}) and their eigenvalues are given by
	\[\label{eigenvaluesJ}
	\textbf{J}^2 Y_{\mu j m}(\theta,\phi)=
	-j(j+1)Y_{\mu j m}(\theta,\phi),
	\]
	where $\mu$ is defined in \eqref{eq:con}.
	The monopole harmonics are given by the hypergeometric function, which is regular once the following quantization condition is imposed
	\[\label{eq:quantCond}
	\mu &= -n \in \mathbb{Z}, \\
	j &= \left| \mu \right|,\left| \mu \right|+1,\left| \mu \right|+2,\dots,\\
	m&= -j,-j+1,\dots,+j.
	\]
	We refer the reader to appendix \ref{app:monopoleHarmonics} for more details.
	The wavefunction therefore takes the separable form
	\begin{equation}
		\Psi = \psi_{njm}e^{-i\omega t}
		= R_{nj}(r)Y_{\mu j m}(\theta,\phi)e^{-i\omega t},
	\end{equation}
	where $R_{nj}(r)$ is determined by the radial equation.

	\subsection{The Hydrogen Atom}
	
	For generic black holes, the radial equation takes the Heun form with four regular singular points. It is not integrable nor it admits a simple analytic solution.
	However, at the SD point $M=N$ it reduces to the confluent hypergeometric equation, which has only two singularities at $r=M$ and $r\to \infty$ and admits a polynomial solution in terms of elementary functions.
	To see this, let us define the following variable
	\begin{equation}
		\rho \equiv \omega (r-M)
	\end{equation}
	and the radial wavefunction
	\begin{equation}
		u_{nj} \equiv (r-M) R_{nj},
	\end{equation}
	in terms of which the radial equation takes the form
	\begin{equation}
		u''_{nj}(\rho)=\left( 1-\frac{\rho_0}{\rho}+\frac{j(j+1)}{\rho^2}  \right) u_{nj}(\rho),\label{eq:HydAtom}
	\end{equation}
	where $\rho_0=-2\mu$. Remarkably, the radial wave equation on the SD black hole \eqref{eq:HydAtom} is \emph{exactly} of the same form as the radial wave equation of the Hydrogen atom (see \cite{griffiths:quantum} for example), with its origin placed at the horizon. Its solutions are given in terms of the associated Laguerre polynomials
	\begin{equation}\label{LaguerreSol}
		u_{nj}(\rho)=\rho^{j+1}e^{-\rho}L^{2j+1}_{n-j-1}(2\rho)
	\end{equation}
	and the complete wavefunction is given by
	\[  \label{intergrableWave}
	\psi_{njm}(y,\theta,\phi) &=\sqrt{\left(\frac{n}{M} \right)^3\frac{(n-j-1)!}{2n ((n+j)!)^3}}\quad\times 
	\\ &\mspace{-48mu}
	e^{-\frac{n}{2M}y}\left(\frac{n}{M}y\right)^j L^{2j+1}_{n-j-1}\left(\frac{n}{M}y\right)
	Y_{njm}(\theta,\phi),
	\]
	where $y\equiv r-n$. The solution in terms of the Laguerre polynomials \eqref{LaguerreSol} is constructed by imposing regularity at the horizon.
	The quantization condition \eqref{eq:quantCond} ensures the regularity of the radial solution (in addition to the regularity of the angular part) and it implies the quantization of the spectrum
	\begin{equation}
		\omega_n = \frac{n}{2M}.
		\label{integrableSpectrum}
	\end{equation}
	The analogue of Bohr's radius is given by $a_0 \rightarrow \frac{2M}{\mu ^2}$.

	\subsection{The Fine Structure of Black Holes}
	
	Hidden symmetries are known to be approximate for generic (i.e. non-extremal) black holes \cite{Castro:2010fd}. Motivated by this we now leverage the integrable structure of the SD point to provide a systematic treatment away from it. We will use perturbation theory to compute corrections to the spectrum. The structure of these corrections is similar to fine structure corrections of the Hydrogen atom and, as we will see shortly, breaks the high degeneracy of the unperturbed spectrum, signaling the underlying symmetry breaking pattern.
	
	We define the perturbation parameter
	\begin{equation}
		\alpha\equiv M-N,
	\end{equation}
	which measures the distance from the SD point, and we use the following rescaling of the wavefunction
	\begin{equation}
		\tilde{\psi}_{njm}\equiv \sqrt{1-2\alpha \left(\frac{1}{y}+\frac{M}{y^2}\right)} \psi_{njm},
	\end{equation}
	for convenience. In terms of the rescaled wavefunction, the wave equation takes the form
	\begin{equation}
		H \tilde{\psi}_{njm} = \left(1+\frac{2M}{y}\right)^2 \omega ^2 \tilde{\psi}_{njm}
		\label{HamiltonianEq}
	\end{equation}
	with the operator
	\begin{equation}
		H=H_0 +\alpha V_1 +\alpha^2 V_2
	\end{equation}
	playing the role of the Hamilonian. Here
	\begin{equation}
		H_0 = \frac{1}{y^2}\partial_y y^2 \partial_y-\frac{\textbf{L}^2}{y^2}
	\end{equation}
	is the Hamiltonian at the SD point and $\textbf{L}$ is the orbital momentum operator. Note that $\textbf{L}$ does not describe a symmetry generator, but $\textbf{L}^2$ is related to the Casimir $\textbf{J}^2$ by \eqref{CasimirsRelation}.
	The potential terms are given by
	\begin{align}
		V_1 &= - 4 \frac{y+M}{y^2}H_0 +\frac{2M}{y^4}-2 \frac{y+M}{y^4} \textbf{L}^2, \\
		V_2 &= + 4 \left(\frac{y+M}{y^2}\right)^2 H_0+\frac{1}{y^4} +4 \frac{(y+M)^2}{y^6} \textbf{L}^2.
	\end{align}
	Let us emphasize that so far we have not used any approximation and that \eqref{HamiltonianEq} is exact. It is quite remarkable that $H$ is exactly quadratic in the perturbation parameter $\alpha$, which is the analogue of the fine structure constant.

	We can now use perturbation theory to compute the correction to the energy levels order by order in $\alpha$
	\[
	\omega_n=\omega_n^{(0)}+\alpha \omega_n^{(1)}+\alpha^2 \omega_n^{(2)}+\dots .
	\]
	The leading order term $\omega_n^{(0)}$ is given by \eqref{integrableSpectrum}. The wavefunction also gets corrected, but these corrections do not contribute to the leading correction to the spectrum $\omega_n^{(1)}$ and we will not consider them here.
	
	Using the wave equation \eqref{HamiltonianEq} we can now compute the leading order correction to the spectrum in terms of the deformation potential
	\[
	\left( \omega_n^{(1)} \right)^2  &= \frac{\braket{\tilde{\psi}_{njm}^0|V_1|\tilde{\psi}_{njm}^0}}{\braket{\tilde{\psi}_{njm}^0|\left(1+\frac{2M}{y}\right)^2|\tilde{\psi}_{njm}^0}}
	\\
	&\approx \frac{\braket{\psi_{njm}^0|V_1|\psi_{njm}^0}}{\braket{\psi_{njm}^0|\left(1+\frac{2M}{y}\right)^2|\psi_{njm}^0}},\\
	&
	\]
	where the second equality holds because we are working to leading order in perturbation theory and $\psi_{njm}^0$ is the unperturbed wavefunction \eqref{intergrableWave}.
	The denominator is given solely in terms of expectation values of inverse powers of $y$ between the unperturbed states
	\[
	\braket{\psi_{njm}^0|\left(1+\frac{2M}{y}\right)^2|\psi_{njm}^0} = 
	1 &+4M \Big<\frac{1}{y} \Big>  \\ 
	&+4M^2 \Big<\frac{1}{y^2} \Big>.
	\]
	The numerator is given by
	\[
	\braket{\psi_{njm}^0|V_1|\psi_{njm}^0}  =&
	-4\left( \omega_n^{(0)} \right)^2  \Big<\frac{1}{y} \Big>  \\
	& -20 M \left( \omega_n^{(0)} \right)^2  \Big<\frac{1}{y^2} \Big>  \\   
	& \mspace{-96mu}  -2\left( l(l+1) +16 M^2 \left( \omega_n^{(0)} \right)^2 \right) \Big<\frac{1}{y^3} \Big> \\
	& \mspace{-96mu} +2M\left( 1-8 M^2 \left( \omega_n^{(0)} \right)^2 -l(l+1) \right)\Big<\frac{1}{y^4} \Big>.
	\]
	where $l(l+1)$ is the eigenvalue of the operator $\textbf{L}^2$, which is related to the $\textbf{J}^2$ eigenvalue by \eqref{CasimirsRelation}.
	Using the known expressions for the expectation values of inverse powers of $y$ (see appendix \ref{app:expectationValues}), and the exact result for the unperturbed energy levels, we arrive at the final result
	\begin{widetext}
		\[
		\left( \omega_n^{(1)} \right)^2 = 
		\frac{n^2}{2M^3}
		\left(
		\frac{
			3n^5
			-2n^3(5j(j+1)-3)
			+nj(j+1)(15j(j+1)-11)
		}
		{
			8(j-\frac{1}{2})j(j+1)(j+\frac{3}{2})(j-n+\frac{1}{2})
		}
		-
		\frac{j+\frac{1}{2}}{j-n+\frac{1}{2}}
		\right).
		\]
	\end{widetext}
	for the leading order correction to the spectrum.
	We see that the degeneracy of the spectrum is lifted and that the energy levels split according to their angular momentum quantum number $j$.

	\section{Discussion}

	Recent years have seen a surge of interest in black hole physics and a growing amount of work on the subject, on both theoretical and experimental fronts. However, precision methods and analytical tools are still very limited. To address this challenge we have put forward two defining questions that motivated our work: is there and integrable Black Hole model and if so, how well does it approximate real world Kerr Black Holes?
	
	We have seen that the SD Black Hole constitutes such an integrable model, in the sense that the wave equation on this background is solvable and its normal modes can be computed exactly. The approximate near-horizon symmetries of the Kerr Black Hole become exact symmetries of the SD Black Hole. Furthermore, we have seen that correction to the spectrum are described by a Hamiltonian which is \emph{exactly} quadratic in the deformation parameter. In this sense the SD Black Hole provides an excellent approximation to real world Kerr Black Holes.
	
	A related work has recently appeared in \cite{Adamo:2023fbj}, where an exact solution of the wave equation on the SD background was studied, albeit with different boundary conditions. The solution of \cite{Adamo:2023fbj} is expressed in terms of momentum eigenstates and reduces to a plane wave in the flat space limit.

	\begin{acknowledgments}
		We thank Tim Adamo, Mikhail Ivanov, Joon-Hwi Kim, Lionel Mason, Noah Miller, Richard Myers and Andrew Strominger for stimulating discussions. We thank Atul Sharma and Chiara Toldo for collaboration on related projects. AG is supported by the Black Hole Initiative and the Society of Fellows at Harvard University, as well as the Department of Energy under grant
		DE-SC0007870.
		UK is supported by the Center for Mathematical Sciences and Applications at Harvard University.
	\end{acknowledgments}
	
	\appendix
	
	\section{Near-Zone Symmetries}\label{app:nearzonesym}

	The near-zone wave equation on the Kerr Taub-NUT background is 
	\begin{equation}
		\Box_{near}\psi:=\left(L^2-\textbf{J}^2 \right)\psi\,,
	\end{equation}
	where
	\begin{align}
		L^2=&\frac{\bar{L}_{+}\bar{L}_{-}+\bar{L}_{+}\bar{L}_{-}}{2}-\bar{L}_{0}^{2}\,, \nonumber\\
		=& \frac{L_{+}L_{-}+L_{+}L_{-}}{2}-L_{0}^{2}\,.
	\end{align}
	In Lorentzian signature, the $SL(2,\mathbb{R})\times\overline{SL(2,\mathbb{R})}$ symmetry generators are given by
	\[
	L_+ &= i \, e^{-2\pi T_R \phi}
	\Big(
	+\sqrt{\Delta} \, \partial_r +  \frac{1}{2\pi T_R} \frac{r-M}{\sqrt{\Delta}} \partial_{\phi}
	\\
	&\qquad\qquad\qquad\qquad\qquad +\frac{2T_L}{T_R} \frac{M \, r -a^2}{\sqrt{\Delta}}\partial_t
	\Big),
	\\ 
	L_0 &= \frac{i}{2\pi T_R} \partial_{\phi}
	+2iM\frac{T_L}{T_R}\partial_t,
	\\  
	L_- &=i \, e^{+2\pi T_R \phi}
	\Big(
	-\sqrt{\Delta} \, \partial_r +  \frac{1}{2\pi T_R} \frac{r-M}{\sqrt{\Delta}} \partial_{\phi}
	\\
	&\qquad\qquad\qquad\qquad\qquad  +\frac{2T_L}{T_R} \frac{M \, r -a^2}{\sqrt{\Delta}}\partial_t
	\Big),
	\]
	and
	\begin{align}
		\bar{L}_+ &= i \, e^{-2\pi T_L \phi+\frac{t}{2M}}
		\Big(
		+\sqrt{\Delta} \, \partial_r - \frac{a}{\sqrt{\Delta}}\partial_{\phi}
		-2M \frac{r}{\sqrt{\Delta}}\partial_t
		\Big),
		\nonumber \\ \nonumber \\
		\bar{L}_0 &=     -2iM \partial_t,
		\\
		\nonumber \\
		\bar{L}_- &=i \, e^{+2\pi T_L \phi-\frac{t}{2M}}
		\Big(
		-\sqrt{\Delta} \, \partial_r - \frac{a}{\sqrt{\Delta}}\partial_{\phi}
		-2M \frac{r}{\sqrt{\Delta}}\partial_t
		\Big).
		\nonumber
	\end{align}
	
	As explained in the main text, these generators are not regular under $\phi \to \phi +2\pi$. Furthermore the expressions for $T_L,T_R$ diverge in the $a\to 0$ limit. A particular modification of the symmetry which avoids this, dubbed Love symmetry, was proposed in \cite{Charalambous:2021mea}. The generators read
	\begin{align}
		\bar{L}_+' &= i \, e^{+\frac{t}{\beta}}
		\Big(
		+\sqrt{\Delta} \, \partial_r - \frac{a}{\sqrt{\Delta}}\partial_{\phi}
		-\beta \, \partial_r \sqrt{\Delta}\partial_t
		\Big),
		\nonumber \\ \nonumber \\
		\bar{L}_0' &=     -i \beta \partial_t,
		\\
		\nonumber \\
		\bar{L}_-' &=i \, e^{-\frac{t}{\beta}}
		\Big(
		-\sqrt{\Delta} \, \partial_r - \frac{a}{\sqrt{\Delta}}\partial_{\phi}
		-\beta \, \partial_r \sqrt{\Delta}\partial_t
		\Big),
		\nonumber
	\end{align}
	where $\beta=2\frac{r_+^2-a^2-N^2}{r_+-r_-}$. At the SD point $T_L \to 0 $ and $\beta \to 2M$ such that $\bar{L}_n'\approx L_n $ up to a term $M\partial_t$ which is irrelevant for $M\omega \ll M/r$ \cite{Castro:2010fd}.

	\section{Near Extremal limits}\label{sec:nenhks}
	
	In this appendix we will study the near-NHEK region of the Kerr black
	hole in the presence of a NUT charge. 
	
	It is first convenient to write down the metric \eqref{eq:sdfs} in Plebanski coordinates,
	namely
	\begin{align}
		q & =r,\,\,\,\,p=N+a\cosh\theta\,,\nonumber \\
		\tau & =t+\frac{a^{2}+N^{2}}{a}\phi,\,\,\,\sigma=\frac{\phi}{a}\,.\label{eq:fds}
	\end{align}
	We obtain
	\begin{align}
		ds^{2} & =(p^{2}-q^{2})\left(\frac{dp^{2}}{\Delta_{p}}-\frac{dq^{2}}{\Delta_{q}}\right)+\frac{1}{p^{2}-q^{2}}\times\nonumber \\
		& \left(\Delta_{p}(d\tau-q^{2}d\sigma)^{2}-\Delta_{q}(d\tau-p^{2}d\sigma)^{2}\right)\,,
	\end{align}
	with
	\begin{equation}
		\Delta_{x}=x^{2}-2M_{x}x+N^{2}-a^{2}\,\,,x=p,q\,,
	\end{equation}
	where $M_{p}=N$ and $M_{q}=M$. The extremal case occurs when the
	zeros of $\Delta_{q}$ coincide, namely $a^{2}=N^{2}-M^{2}$.
	
	Recall that the Hawking temperature and horizon velocity are given by
	\begin{equation}
		T_{H}=\frac{r_{+}-r_{-}}{4\pi(r_{+}^{2}-N^{2}-a^{2})},\quad \Omega_{+}=\frac{a}{r_{+}^{2}-N^{2}-a^{2}}.
	\end{equation}
	As defined in the text $T_{R}=T_{H}/\Omega_{+}$ vanishes in the extremal
	limit. Motivated by the analysis of \cite{Bredberg:2009pv}, to obtain a near extremal temperature we introduce a redshift
	factor $a/\lambda M$
	
	\begin{equation}
		\frac{\mathcal{T}}{2\pi}=\frac{T_{R}}{\lambda}\times\frac{a}{M}=\frac{r_{+}-r_{-}}{4\pi\lambda M}
	\end{equation}
	and take $\lambda\to0$ while keeping the horizon temp $\mathcal{T}$
	fixed, where $\mathcal{T}=0$ corresponds to the exact extremal BH.
	Using \eqref{eq:rpms}, this leads to
	
	\begin{align}
		a & =\sqrt{N^{2}-M^{2}+(\mathcal{T}M\lambda)^{2}}\nonumber \\
		& \approx\sqrt{N^{2}-M^{2}}+\frac{(\mathcal{T}M\lambda)^{2}}{2\sqrt{N^{2}-M^{2}}}
	\end{align}
	(note that we obtain $T_{H}\approx\frac{\lambda M\mathcal{T}}{4\pi(M^{2}-N^{2})}\to0$), away from the SD point $N\neq M$. At the strict SD point we find instead $a=\mathcal{T M \lambda}$, hence $\mathcal{T}$ is a rotation parameter. Moreover, since $a$ is pure gauge at the SD point, the limit $\lambda \to 0$ is completely irrelevant! thus suggesting that the SD point is self-similar (and that $\mathcal{T}$ can be shifted by a diffeomorphism as will be argued in general).
	
	As $\lambda\to0$ we further approach the horizon via
	
	\begin{equation}
		r'=\frac{q-M}{\lambda M}\,,t'=\frac{\lambda\tau}{M}\,,\phi'=\frac{M\sigma}{2}-\frac{\tau}{2M}.
	\end{equation}
	The result is 
	\begin{align}
		ds_{\textrm{NNHEK}}^{2} & =(p^{2}-M^{2})\left(\frac{dp^{2}}{\tilde{\Delta}}+(r'^{2}-\mathcal{T}^{2})dt'^{2}
		+\right.\nonumber \\
		&\left. \qquad \quad -\frac{dr'^{2}}{r'^{2}-\mathcal{T}^{2}}
		f(p)(r'dt'+d\phi')^{2}\right),\label{eq:nhekm}
	\end{align}
	where $\tilde{\Delta}$ and $f(p)$ are given in the main text. As it turns out, the temperature $\mathcal{T}$ can be removed locally
	by a change of coordinates, see e.g. \cite{Bredberg:2009pv,Anninos:2008fx}. Indeed, the metric \eqref{eq:nhekm} coincides with \eqref{ec:nnsk} if we take the basis of one-forms to be
	\begin{align}
		\sigma_{+} & =e^{\frac{i\phi'}{2}}\left((r'-\mathcal{T})dt'-\frac{dr'}{r'+\mathcal{T}}\right),\nonumber \\
		\sigma_{-} & =e^{-\frac{i\phi'}{2}}\left((r'+\mathcal{T})dt'+\frac{dr'}{r'-\mathcal{T}}\right),\nonumber \\
		\sigma_{0} & =r'dt'+d\phi'
	\end{align}
	and identify $p=\tilde{r}$. In particular the constant $\theta$ leaves, namely $dp=0$, are warped $AdS_{3}$ `black holes' \cite{Anninos:2008fx}. Just as in unwarped $AdS_{3}$ (BTZ black hole), different values
	of $\mathcal{T}$ correspond to different
	coordinate patches, where $\mathcal{T}=1$ being a global patch. Indeed, if $\mathcal{T}\neq 0$ we can remove it locally via the transformation
	\begin{align}
		p & \to \tilde{r},\,\,\,\phi' \to \tilde{t}/2M, \nonumber \\
		r' & \to\mathcal{T}\cosh\tilde{\theta},\,\,\,t'\to\tilde{\phi}/\mathcal{T},
	\end{align}
	that bring the metric into the form
	\[
	ds^2_{\textrm{NNHEK}}&=\frac{\tilde{\Delta}}{\tilde{r}^{2}{-}M^{2}}(d\tilde{t}{-}2M\cosh\tilde{\theta})^{2}{+}\\
	&+(\tilde{r}^{2}{-}M^{2})(\frac{d\tilde{r}^{2}}{\tilde{\Delta}}{-}d\tilde{\theta}^{2}{-}\sinh\tilde{\theta}d\tilde{\phi}^{2}),
	\]
	which is nothing but the static instance ($a=0$) of \eqref{eq:sdfs}, with $M$ and $N$ exchanged (see \cite{Crawley:2021auj}).
	This confirms that for any value
	of the spin $a$, the (near)NHEK limit at the self-dual point is nothing
	but a diffeomorphism. Finally, note that $\mathcal{T}$
	is simply generated by boosting with $L_{0}$ on the $t',r'$ plane, where $\mathcal{T}=0$ corresponds to infinite boost.

	\section{Monopole Harmonics}\label{app:monopoleHarmonics}
	
	The Taub-NUT metric has four Killing vectors \cite{Zimmerman:1989kv}. One of them corresponds to time translations and the rest to rotational invariance. The angular momentum operator is decomposed into the orbital and spin parts
	\begin{equation}
		\textbf{J} = \textbf{L}+\textbf{S},
	\end{equation}
	where
	\[
	\textbf{L} &= \textbf{r}\times \left( \textbf{p}+ 2 i \textbf{A} \partial_t \right)
	\quad \text{with} \quad \textbf{A} = N \frac{\cosh\theta}{r \sinh \theta} \hat{\phi},
	\\
	\textbf{S} &= -  2iN\,  \hat{r} \,  \partial_t.
	\]
	The Casimir operator of the spherical symmetry
	\[\label{angularSquare}
	\textbf{J}^2 = &-\frac{1}{\sinh\theta}\partial_{\theta}\sinh\theta\partial_{\theta}
	+4N^2 \partial_t^2
	\\
	& -\frac{1}{\sinh^2\theta}\left(\partial_{\phi}+2N(1+\cosh\theta)\partial_t\right)^2
	\]
	is related to orbital angular momentum squared by
	\begin{equation}
		\textbf{J}^2 = \textbf{L}^2 + \mu^2,
		\label{CasimirsRelation}
	\end{equation}
	where $\mu$ is defined in \eqref{eq:con}.
	
	The eigenfunctions of $\textbf{J}^2$ are the \emph{monopole harmonics} \cite{Wu:1976ge,Shnir:2005vvi}, which are given by
	\[
	Y_{\mu j m}(x,\phi) &= 2^m \sqrt{\frac{2j+1}{4\pi}\frac{(j-m)!(j+m)!}{(j-\mu)!(j+\mu)!}} \, \times  \\
	&\mspace{-64mu} e^{i(\mu+m)\phi}(1+x)^{-\frac{\mu+m}{2}}(1-x)^{-\frac{\mu-m}{2}} \, \times  \\
	&\mspace{-64mu}{}_2F_1\left( -j-\mu,1+j-\mu;1+m-\mu;\frac{1-x}{2}\right),
	\]
	where $x=\cosh \theta$, and its eigenvalues are given in \eqref{eigenvaluesJ}. Let us also note that the eigenvalues of the orbital angular momentum operator $\textbf{L}$ are related to those of $\textbf{J}$ by
	\begin{equation}
		j(j+1)=l(l+1)+\mu^2.
	\end{equation}

	\section{Hydrogen Atom Expectation Values}\label{app:expectationValues}

	The expectation values of inverse powers of $r$ between Hydrogen atom states can be read from the book of Bethe and Salpeter \cite{Bethe:1957ncq} and are given by
	\[
	\Big<\frac{1}{y}\Big> &= \frac{1}{n^2 a_0}, \\
	\Big<\frac{1}{y^2}\Big> &= \frac{1}{(j+\frac{1}{2})n^3 a_0^2},\\
	\Big<\frac{1}{y^3}\Big> &= \frac{1}{j(j+\frac{1}{2})(j+1)n^3a_0^3},\\
	\Big<\frac{1}{y^4}\Big> &= \frac{
		3n^2 - j(j+1)
	}{
		2n^5a_0^4 (j-\frac{1}{2})j(j+\frac{1}{2})(j+1)(j+\frac{3}{2})
	}.
	\]
	Here $a_0$ is the Bohr radius, which in the case of the SD black hole is given by $\frac{2M}{\mu^2}=\frac{2M}{n^2}$.

	
	\bibliography{apssamp}

\providecommand{\noopsort}[1]{}\providecommand{\singleletter}[1]{#1}%
\begin{thebibliography}{51}%
\makeatletter
\providecommand \@ifxundefined [1]{%
 \@ifx{#1\undefined}
}%
\providecommand \@ifnum [1]{%
 \ifnum #1\expandafter \@firstoftwo
 \else \expandafter \@secondoftwo
 \fi
}%
\providecommand \@ifx [1]{%
 \ifx #1\expandafter \@firstoftwo
 \else \expandafter \@secondoftwo
 \fi
}%
\providecommand \natexlab [1]{#1}%
\providecommand \enquote  [1]{``#1''}%
\providecommand \bibnamefont  [1]{#1}%
\providecommand \bibfnamefont [1]{#1}%
\providecommand \citenamefont [1]{#1}%
\providecommand \href@noop [0]{\@secondoftwo}%
\providecommand \href [0]{\begingroup \@sanitize@url \@href}%
\providecommand \@href[1]{\@@startlink{#1}\@@href}%
\providecommand \@@href[1]{\endgroup#1\@@endlink}%
\providecommand \@sanitize@url [0]{\catcode `\\12\catcode `\$12\catcode
  `\&12\catcode `\#12\catcode `\^12\catcode `\_12\catcode `\%12\relax}%
\providecommand \@@startlink[1]{}%
\providecommand \@@endlink[0]{}%
\providecommand \url  [0]{\begingroup\@sanitize@url \@url }%
\providecommand \@url [1]{\endgroup\@href {#1}{\urlprefix }}%
\providecommand \urlprefix  [0]{URL }%
\providecommand \Eprint [0]{\href }%
\providecommand \doibase [0]{http://dx.doi.org/}%
\providecommand \selectlanguage [0]{\@gobble}%
\providecommand \bibinfo  [0]{\@secondoftwo}%
\providecommand \bibfield  [0]{\@secondoftwo}%
\providecommand \translation [1]{[#1]}%
\providecommand \BibitemOpen [0]{}%
\providecommand \bibitemStop [0]{}%
\providecommand \bibitemNoStop [0]{.\EOS\space}%
\providecommand \EOS [0]{\spacefactor3000\relax}%
\providecommand \BibitemShut  [1]{\csname bibitem#1\endcsname}%
\let\auto@bib@innerbib\@empty
\bibitem [{\citenamefont {Dijkgraaf}(2019)}]{dijkgraaf2019}%
  \BibitemOpen
  \bibfield  {author} {\bibinfo {author} {\bibfnamefont {R.}~\bibnamefont
  {Dijkgraaf}},\ }\href@noop {} {\enquote {\bibinfo {title} {The black hole is
  the atom of the 21st century},}\ }\bibinfo {howpublished}
  {\url{https://www.ias.edu/ideas/dijkgraaf-EHT-black-hole}} (\bibinfo {year}
  {2019})\BibitemShut {NoStop}%
\bibitem [{\citenamefont {Abbott}\ \emph {et~al.}(2016)\citenamefont {Abbott}
  \emph {et~al.}}]{LIGOScientific:2016lio}%
  \BibitemOpen
  \bibfield  {author} {\bibinfo {author} {\bibfnamefont {B.~P.}\ \bibnamefont
  {Abbott}} \emph {et~al.} (\bibinfo {collaboration} {LIGO Scientific,
  Virgo}),\ }\href {\doibase 10.1103/PhysRevLett.116.221101} {\bibfield
  {journal} {\bibinfo  {journal} {Phys. Rev. Lett.}\ }\textbf {\bibinfo
  {volume} {116}},\ \bibinfo {pages} {221101} (\bibinfo {year} {2016})},\
  \bibinfo {note} {[Erratum: Phys.Rev.Lett. 121, 129902 (2018)]},\ \Eprint
  {http://arxiv.org/abs/1602.03841} {arXiv:1602.03841 [gr-qc]} \BibitemShut
  {NoStop}%
\bibitem [{\citenamefont {Akiyama}\ \emph {et~al.}(2019)\citenamefont {Akiyama}
  \emph {et~al.}}]{EventHorizonTelescope:2019dse}%
  \BibitemOpen
  \bibfield  {author} {\bibinfo {author} {\bibfnamefont {K.}~\bibnamefont
  {Akiyama}} \emph {et~al.} (\bibinfo {collaboration} {Event Horizon
  Telescope}),\ }\href {\doibase 10.3847/2041-8213/ab0ec7} {\bibfield
  {journal} {\bibinfo  {journal} {Astrophys. J. Lett.}\ }\textbf {\bibinfo
  {volume} {875}},\ \bibinfo {pages} {L1} (\bibinfo {year} {2019})},\ \Eprint
  {http://arxiv.org/abs/1906.11238} {arXiv:1906.11238 [astro-ph.GA]}
  \BibitemShut {NoStop}%
\bibitem [{\citenamefont {Teukolsky}(1972)}]{Teukolsky:1972my}%
  \BibitemOpen
  \bibfield  {author} {\bibinfo {author} {\bibfnamefont {S.~A.}\ \bibnamefont
  {Teukolsky}},\ }\href {\doibase 10.1103/PhysRevLett.29.1114} {\bibfield
  {journal} {\bibinfo  {journal} {Phys. Rev. Lett.}\ }\textbf {\bibinfo
  {volume} {29}},\ \bibinfo {pages} {1114} (\bibinfo {year}
  {1972})}\BibitemShut {NoStop}%
\bibitem [{\citenamefont {Teukolsky}(1973)}]{Teukolsky:1973ha}%
  \BibitemOpen
  \bibfield  {author} {\bibinfo {author} {\bibfnamefont {S.~A.}\ \bibnamefont
  {Teukolsky}},\ }\href {\doibase 10.1086/152444} {\bibfield  {journal}
  {\bibinfo  {journal} {Astrophys. J.}\ }\textbf {\bibinfo {volume} {185}},\
  \bibinfo {pages} {635} (\bibinfo {year} {1973})}\BibitemShut {NoStop}%
\bibitem [{\citenamefont {Futterman}\ \emph {et~al.}(1988)\citenamefont
  {Futterman}, \citenamefont {Handler},\ and\ \citenamefont
  {Matzner}}]{futterman_handler_matzner_1988}%
  \BibitemOpen
  \bibfield  {author} {\bibinfo {author} {\bibfnamefont {J.~A.~H.}\
  \bibnamefont {Futterman}}, \bibinfo {author} {\bibfnamefont {F.~A.}\
  \bibnamefont {Handler}}, \ and\ \bibinfo {author} {\bibfnamefont {R.~A.}\
  \bibnamefont {Matzner}},\ }\href {\doibase 10.1017/CBO9780511735615} {\emph
  {\bibinfo {title} {Scattering from Black Holes}}},\ Cambridge Monographs on
  Mathematical Physics\ (\bibinfo  {publisher} {Cambridge University Press},\
  \bibinfo {year} {1988})\BibitemShut {NoStop}%
\bibitem [{\citenamefont {Yang}\ \emph {et~al.}(2012)\citenamefont {Yang},
  \citenamefont {Nichols}, \citenamefont {Zhang}, \citenamefont {Zimmerman},
  \citenamefont {Zhang},\ and\ \citenamefont {Chen}}]{Yang:2012he}%
  \BibitemOpen
  \bibfield  {author} {\bibinfo {author} {\bibfnamefont {H.}~\bibnamefont
  {Yang}}, \bibinfo {author} {\bibfnamefont {D.~A.}\ \bibnamefont {Nichols}},
  \bibinfo {author} {\bibfnamefont {F.}~\bibnamefont {Zhang}}, \bibinfo
  {author} {\bibfnamefont {A.}~\bibnamefont {Zimmerman}}, \bibinfo {author}
  {\bibfnamefont {Z.}~\bibnamefont {Zhang}}, \ and\ \bibinfo {author}
  {\bibfnamefont {Y.}~\bibnamefont {Chen}},\ }\href {\doibase
  10.1103/PhysRevD.86.104006} {\bibfield  {journal} {\bibinfo  {journal} {Phys.
  Rev. D}\ }\textbf {\bibinfo {volume} {86}},\ \bibinfo {pages} {104006}
  (\bibinfo {year} {2012})},\ \Eprint {http://arxiv.org/abs/1207.4253}
  {arXiv:1207.4253 [gr-qc]} \BibitemShut {NoStop}%
\bibitem [{\citenamefont {Berti}\ \emph {et~al.}(2009)\citenamefont {Berti},
  \citenamefont {Cardoso},\ and\ \citenamefont {Starinets}}]{Berti:2009kk}%
  \BibitemOpen
  \bibfield  {author} {\bibinfo {author} {\bibfnamefont {E.}~\bibnamefont
  {Berti}}, \bibinfo {author} {\bibfnamefont {V.}~\bibnamefont {Cardoso}}, \
  and\ \bibinfo {author} {\bibfnamefont {A.~O.}\ \bibnamefont {Starinets}},\
  }\href {\doibase 10.1088/0264-9381/26/16/163001} {\bibfield  {journal}
  {\bibinfo  {journal} {Class. Quant. Grav.}\ }\textbf {\bibinfo {volume}
  {26}},\ \bibinfo {pages} {163001} (\bibinfo {year} {2009})},\ \Eprint
  {http://arxiv.org/abs/0905.2975} {arXiv:0905.2975 [gr-qc]} \BibitemShut
  {NoStop}%
\bibitem [{\citenamefont {Pauli}(1926)}]{Pauli:1926qpj}%
  \BibitemOpen
  \bibfield  {author} {\bibinfo {author} {\bibfnamefont {W.}~\bibnamefont
  {Pauli}},\ }\href {\doibase 10.1007/BF01450175} {\bibfield  {journal}
  {\bibinfo  {journal} {Z. Phys.}\ }\textbf {\bibinfo {volume} {36}},\ \bibinfo
  {pages} {336} (\bibinfo {year} {1926})}\BibitemShut {NoStop}%
\bibitem [{\citenamefont {Caron-Huot}\ and\ \citenamefont
  {Henn}(2014)}]{Caron-Huot:2014gia}%
  \BibitemOpen
  \bibfield  {author} {\bibinfo {author} {\bibfnamefont {S.}~\bibnamefont
  {Caron-Huot}}\ and\ \bibinfo {author} {\bibfnamefont {J.~M.}\ \bibnamefont
  {Henn}},\ }\href {\doibase 10.1103/PhysRevLett.113.161601} {\bibfield
  {journal} {\bibinfo  {journal} {Phys. Rev. Lett.}\ }\textbf {\bibinfo
  {volume} {113}},\ \bibinfo {pages} {161601} (\bibinfo {year} {2014})},\
  \Eprint {http://arxiv.org/abs/1408.0296} {arXiv:1408.0296 [hep-th]}
  \BibitemShut {NoStop}%
\bibitem [{\citenamefont {Castro}\ \emph {et~al.}(2010)\citenamefont {Castro},
  \citenamefont {Maloney},\ and\ \citenamefont {Strominger}}]{Castro:2010fd}%
  \BibitemOpen
  \bibfield  {author} {\bibinfo {author} {\bibfnamefont {A.}~\bibnamefont
  {Castro}}, \bibinfo {author} {\bibfnamefont {A.}~\bibnamefont {Maloney}}, \
  and\ \bibinfo {author} {\bibfnamefont {A.}~\bibnamefont {Strominger}},\
  }\href {\doibase 10.1103/PhysRevD.82.024008} {\bibfield  {journal} {\bibinfo
  {journal} {Phys. Rev. D}\ }\textbf {\bibinfo {volume} {82}},\ \bibinfo
  {pages} {024008} (\bibinfo {year} {2010})},\ \Eprint
  {http://arxiv.org/abs/1004.0996} {arXiv:1004.0996 [hep-th]} \BibitemShut
  {NoStop}%
\bibitem [{\citenamefont {Guica}\ \emph {et~al.}(2009)\citenamefont {Guica},
  \citenamefont {Hartman}, \citenamefont {Song},\ and\ \citenamefont
  {Strominger}}]{Guica:2008mu}%
  \BibitemOpen
  \bibfield  {author} {\bibinfo {author} {\bibfnamefont {M.}~\bibnamefont
  {Guica}}, \bibinfo {author} {\bibfnamefont {T.}~\bibnamefont {Hartman}},
  \bibinfo {author} {\bibfnamefont {W.}~\bibnamefont {Song}}, \ and\ \bibinfo
  {author} {\bibfnamefont {A.}~\bibnamefont {Strominger}},\ }\href {\doibase
  10.1103/PhysRevD.80.124008} {\bibfield  {journal} {\bibinfo  {journal} {Phys.
  Rev. D}\ }\textbf {\bibinfo {volume} {80}},\ \bibinfo {pages} {124008}
  (\bibinfo {year} {2009})},\ \Eprint {http://arxiv.org/abs/0809.4266}
  {arXiv:0809.4266 [hep-th]} \BibitemShut {NoStop}%
\bibitem [{\citenamefont {Bredberg}\ \emph {et~al.}(2010)\citenamefont
  {Bredberg}, \citenamefont {Hartman}, \citenamefont {Song},\ and\
  \citenamefont {Strominger}}]{Bredberg:2009pv}%
  \BibitemOpen
  \bibfield  {author} {\bibinfo {author} {\bibfnamefont {I.}~\bibnamefont
  {Bredberg}}, \bibinfo {author} {\bibfnamefont {T.}~\bibnamefont {Hartman}},
  \bibinfo {author} {\bibfnamefont {W.}~\bibnamefont {Song}}, \ and\ \bibinfo
  {author} {\bibfnamefont {A.}~\bibnamefont {Strominger}},\ }\href {\doibase
  10.1007/JHEP04(2010)019} {\bibfield  {journal} {\bibinfo  {journal} {JHEP}\
  }\textbf {\bibinfo {volume} {04}},\ \bibinfo {pages} {019} (\bibinfo {year}
  {2010})},\ \Eprint {http://arxiv.org/abs/0907.3477} {arXiv:0907.3477
  [hep-th]} \BibitemShut {NoStop}%
\bibitem [{\citenamefont {Charalambous}\ \emph
  {et~al.}(2021{\natexlab{a}})\citenamefont {Charalambous}, \citenamefont
  {Dubovsky},\ and\ \citenamefont {Ivanov}}]{Charalambous:2021mea}%
  \BibitemOpen
  \bibfield  {author} {\bibinfo {author} {\bibfnamefont {P.}~\bibnamefont
  {Charalambous}}, \bibinfo {author} {\bibfnamefont {S.}~\bibnamefont
  {Dubovsky}}, \ and\ \bibinfo {author} {\bibfnamefont {M.~M.}\ \bibnamefont
  {Ivanov}},\ }\href {\doibase 10.1007/JHEP05(2021)038} {\bibfield  {journal}
  {\bibinfo  {journal} {JHEP}\ }\textbf {\bibinfo {volume} {05}},\ \bibinfo
  {pages} {038} (\bibinfo {year} {2021}{\natexlab{a}})},\ \Eprint
  {http://arxiv.org/abs/2102.08917} {arXiv:2102.08917 [hep-th]} \BibitemShut
  {NoStop}%
\bibitem [{\citenamefont {Charalambous}\ \emph
  {et~al.}(2021{\natexlab{b}})\citenamefont {Charalambous}, \citenamefont
  {Dubovsky},\ and\ \citenamefont {Ivanov}}]{Charalambous:2021kcz}%
  \BibitemOpen
  \bibfield  {author} {\bibinfo {author} {\bibfnamefont {P.}~\bibnamefont
  {Charalambous}}, \bibinfo {author} {\bibfnamefont {S.}~\bibnamefont
  {Dubovsky}}, \ and\ \bibinfo {author} {\bibfnamefont {M.~M.}\ \bibnamefont
  {Ivanov}},\ }\href {\doibase 10.1103/PhysRevLett.127.101101} {\bibfield
  {journal} {\bibinfo  {journal} {Phys. Rev. Lett.}\ }\textbf {\bibinfo
  {volume} {127}},\ \bibinfo {pages} {101101} (\bibinfo {year}
  {2021}{\natexlab{b}})},\ \Eprint {http://arxiv.org/abs/2103.01234}
  {arXiv:2103.01234 [hep-th]} \BibitemShut {NoStop}%
\bibitem [{\citenamefont {Charalambous}\ \emph {et~al.}(2022)\citenamefont
  {Charalambous}, \citenamefont {Dubovsky},\ and\ \citenamefont
  {Ivanov}}]{Charalambous:2022rre}%
  \BibitemOpen
  \bibfield  {author} {\bibinfo {author} {\bibfnamefont {P.}~\bibnamefont
  {Charalambous}}, \bibinfo {author} {\bibfnamefont {S.}~\bibnamefont
  {Dubovsky}}, \ and\ \bibinfo {author} {\bibfnamefont {M.~M.}\ \bibnamefont
  {Ivanov}},\ }\href {\doibase 10.1007/JHEP10(2022)175} {\bibfield  {journal}
  {\bibinfo  {journal} {JHEP}\ }\textbf {\bibinfo {volume} {10}},\ \bibinfo
  {pages} {175} (\bibinfo {year} {2022})},\ \Eprint
  {http://arxiv.org/abs/2209.02091} {arXiv:2209.02091 [hep-th]} \BibitemShut
  {NoStop}%
\bibitem [{\citenamefont {Chia}(2021)}]{Chia:2020yla}%
  \BibitemOpen
  \bibfield  {author} {\bibinfo {author} {\bibfnamefont {H.~S.}\ \bibnamefont
  {Chia}},\ }\href {\doibase 10.1103/PhysRevD.104.024013} {\bibfield  {journal}
  {\bibinfo  {journal} {Phys. Rev. D}\ }\textbf {\bibinfo {volume} {104}},\
  \bibinfo {pages} {024013} (\bibinfo {year} {2021})},\ \Eprint
  {http://arxiv.org/abs/2010.07300} {arXiv:2010.07300 [gr-qc]} \BibitemShut
  {NoStop}%
\bibitem [{\citenamefont {Hui}\ \emph {et~al.}(2022{\natexlab{a}})\citenamefont
  {Hui}, \citenamefont {Joyce}, \citenamefont {Penco}, \citenamefont
  {Santoni},\ and\ \citenamefont {Solomon}}]{Hui:2021vcv}%
  \BibitemOpen
  \bibfield  {author} {\bibinfo {author} {\bibfnamefont {L.}~\bibnamefont
  {Hui}}, \bibinfo {author} {\bibfnamefont {A.}~\bibnamefont {Joyce}}, \bibinfo
  {author} {\bibfnamefont {R.}~\bibnamefont {Penco}}, \bibinfo {author}
  {\bibfnamefont {L.}~\bibnamefont {Santoni}}, \ and\ \bibinfo {author}
  {\bibfnamefont {A.~R.}\ \bibnamefont {Solomon}},\ }\href {\doibase
  10.1088/1475-7516/2022/01/032} {\bibfield  {journal} {\bibinfo  {journal}
  {JCAP}\ }\textbf {\bibinfo {volume} {01}},\ \bibinfo {pages} {032} (\bibinfo
  {year} {2022}{\natexlab{a}})},\ \Eprint {http://arxiv.org/abs/2105.01069}
  {arXiv:2105.01069 [hep-th]} \BibitemShut {NoStop}%
\bibitem [{\citenamefont {Ben~Achour}\ \emph {et~al.}(2022)\citenamefont
  {Ben~Achour}, \citenamefont {Livine}, \citenamefont {Mukohyama},\ and\
  \citenamefont {Uzan}}]{BenAchour:2022uqo}%
  \BibitemOpen
  \bibfield  {author} {\bibinfo {author} {\bibfnamefont {J.}~\bibnamefont
  {Ben~Achour}}, \bibinfo {author} {\bibfnamefont {E.~R.}\ \bibnamefont
  {Livine}}, \bibinfo {author} {\bibfnamefont {S.}~\bibnamefont {Mukohyama}}, \
  and\ \bibinfo {author} {\bibfnamefont {J.-P.}\ \bibnamefont {Uzan}},\ }\href
  {\doibase 10.1007/JHEP07(2022)112} {\bibfield  {journal} {\bibinfo  {journal}
  {JHEP}\ }\textbf {\bibinfo {volume} {07}},\ \bibinfo {pages} {112} (\bibinfo
  {year} {2022})},\ \Eprint {http://arxiv.org/abs/2202.12828} {arXiv:2202.12828
  [gr-qc]} \BibitemShut {NoStop}%
\bibitem [{\citenamefont {Hui}\ \emph {et~al.}(2022{\natexlab{b}})\citenamefont
  {Hui}, \citenamefont {Joyce}, \citenamefont {Penco}, \citenamefont
  {Santoni},\ and\ \citenamefont {Solomon}}]{Hui:2022vbh}%
  \BibitemOpen
  \bibfield  {author} {\bibinfo {author} {\bibfnamefont {L.}~\bibnamefont
  {Hui}}, \bibinfo {author} {\bibfnamefont {A.}~\bibnamefont {Joyce}}, \bibinfo
  {author} {\bibfnamefont {R.}~\bibnamefont {Penco}}, \bibinfo {author}
  {\bibfnamefont {L.}~\bibnamefont {Santoni}}, \ and\ \bibinfo {author}
  {\bibfnamefont {A.~R.}\ \bibnamefont {Solomon}},\ }\href {\doibase
  10.1007/JHEP09(2022)049} {\bibfield  {journal} {\bibinfo  {journal} {JHEP}\
  }\textbf {\bibinfo {volume} {09}},\ \bibinfo {pages} {049} (\bibinfo {year}
  {2022}{\natexlab{b}})},\ \Eprint {http://arxiv.org/abs/2203.08832}
  {arXiv:2203.08832 [hep-th]} \BibitemShut {NoStop}%
\bibitem [{\citenamefont {Huang}\ \emph {et~al.}(2020)\citenamefont {Huang},
  \citenamefont {Kol},\ and\ \citenamefont {O'Connell}}]{Huang:2019cja}%
  \BibitemOpen
  \bibfield  {author} {\bibinfo {author} {\bibfnamefont {Y.-T.}\ \bibnamefont
  {Huang}}, \bibinfo {author} {\bibfnamefont {U.}~\bibnamefont {Kol}}, \ and\
  \bibinfo {author} {\bibfnamefont {D.}~\bibnamefont {O'Connell}},\ }\href
  {\doibase 10.1103/PhysRevD.102.046005} {\bibfield  {journal} {\bibinfo
  {journal} {Phys. Rev. D}\ }\textbf {\bibinfo {volume} {102}},\ \bibinfo
  {pages} {046005} (\bibinfo {year} {2020})},\ \Eprint
  {http://arxiv.org/abs/1911.06318} {arXiv:1911.06318 [hep-th]} \BibitemShut
  {NoStop}%
\bibitem [{\citenamefont {Crawley}\ \emph {et~al.}(2022)\citenamefont
  {Crawley}, \citenamefont {Guevara}, \citenamefont {Miller},\ and\
  \citenamefont {Strominger}}]{Crawley:2021auj}%
  \BibitemOpen
  \bibfield  {author} {\bibinfo {author} {\bibfnamefont {E.}~\bibnamefont
  {Crawley}}, \bibinfo {author} {\bibfnamefont {A.}~\bibnamefont {Guevara}},
  \bibinfo {author} {\bibfnamefont {N.}~\bibnamefont {Miller}}, \ and\ \bibinfo
  {author} {\bibfnamefont {A.}~\bibnamefont {Strominger}},\ }\href {\doibase
  10.1007/JHEP10(2022)135} {\bibfield  {journal} {\bibinfo  {journal} {JHEP}\
  }\textbf {\bibinfo {volume} {10}},\ \bibinfo {pages} {135} (\bibinfo {year}
  {2022})},\ \Eprint {http://arxiv.org/abs/2112.03954} {arXiv:2112.03954
  [hep-th]} \BibitemShut {NoStop}%
\bibitem [{Note1()}]{Note1}%
  \BibitemOpen
  \bibinfo {note} {The symmetry generators in this work can then be
  analytically continued back to Lorentzian signature.}\BibitemShut {Stop}%
\bibitem [{\citenamefont {Castro}\ \emph {et~al.}(2013)\citenamefont {Castro},
  \citenamefont {Lapan}, \citenamefont {Maloney},\ and\ \citenamefont
  {Rodriguez}}]{Castro:2013kea}%
  \BibitemOpen
  \bibfield  {author} {\bibinfo {author} {\bibfnamefont {A.}~\bibnamefont
  {Castro}}, \bibinfo {author} {\bibfnamefont {J.~M.}\ \bibnamefont {Lapan}},
  \bibinfo {author} {\bibfnamefont {A.}~\bibnamefont {Maloney}}, \ and\
  \bibinfo {author} {\bibfnamefont {M.~J.}\ \bibnamefont {Rodriguez}},\ }\href
  {\doibase 10.1103/PhysRevD.88.044003} {\bibfield  {journal} {\bibinfo
  {journal} {Phys. Rev. D}\ }\textbf {\bibinfo {volume} {88}},\ \bibinfo
  {pages} {044003} (\bibinfo {year} {2013})},\ \Eprint
  {http://arxiv.org/abs/1303.0759} {arXiv:1303.0759 [hep-th]} \BibitemShut
  {NoStop}%
\bibitem [{\citenamefont {Perry}\ and\ \citenamefont
  {Rodriguez}(2022)}]{Perry:2022udk}%
  \BibitemOpen
  \bibfield  {author} {\bibinfo {author} {\bibfnamefont {M.~J.}\ \bibnamefont
  {Perry}}\ and\ \bibinfo {author} {\bibfnamefont {M.~J.}\ \bibnamefont
  {Rodriguez}},\ }\href@noop {} {\  (\bibinfo {year} {2022})},\ \Eprint
  {http://arxiv.org/abs/2205.09146} {arXiv:2205.09146 [hep-th]} \BibitemShut
  {NoStop}%
\bibitem [{\citenamefont {Comp\`ere}(2012)}]{Compere:2012jk}%
  \BibitemOpen
  \bibfield  {author} {\bibinfo {author} {\bibfnamefont {G.}~\bibnamefont
  {Comp\`ere}},\ }\href {\doibase 10.1007/s41114-017-0003-2} {\bibfield
  {journal} {\bibinfo  {journal} {Living Rev. Rel.}\ }\textbf {\bibinfo
  {volume} {15}},\ \bibinfo {pages} {11} (\bibinfo {year} {2012})},\ \Eprint
  {http://arxiv.org/abs/1203.3561} {arXiv:1203.3561 [hep-th]} \BibitemShut
  {NoStop}%
\bibitem [{Note2()}]{Note2}%
  \BibitemOpen
  \bibinfo {note} {We can get $T_R=0$ by first taking the extremal limit
  $r_+\to r_{-}$, followed by the SD limit $M\to N$}\BibitemShut {NoStop}%
\bibitem [{\citenamefont {Guevara}(2019)}]{Guevara:2017csg}%
  \BibitemOpen
  \bibfield  {author} {\bibinfo {author} {\bibfnamefont {A.}~\bibnamefont
  {Guevara}},\ }\href {\doibase 10.1007/JHEP04(2019)033} {\bibfield  {journal}
  {\bibinfo  {journal} {JHEP}\ }\textbf {\bibinfo {volume} {04}},\ \bibinfo
  {pages} {033} (\bibinfo {year} {2019})},\ \Eprint
  {http://arxiv.org/abs/1706.02314} {arXiv:1706.02314 [hep-th]} \BibitemShut
  {NoStop}%
\bibitem [{\citenamefont {Guevara}\ \emph {et~al.}(2019)\citenamefont
  {Guevara}, \citenamefont {Ochirov},\ and\ \citenamefont
  {Vines}}]{Guevara:2018wpp}%
  \BibitemOpen
  \bibfield  {author} {\bibinfo {author} {\bibfnamefont {A.}~\bibnamefont
  {Guevara}}, \bibinfo {author} {\bibfnamefont {A.}~\bibnamefont {Ochirov}}, \
  and\ \bibinfo {author} {\bibfnamefont {J.}~\bibnamefont {Vines}},\ }\href
  {\doibase 10.1007/JHEP09(2019)056} {\bibfield  {journal} {\bibinfo  {journal}
  {JHEP}\ }\textbf {\bibinfo {volume} {09}},\ \bibinfo {pages} {056} (\bibinfo
  {year} {2019})},\ \Eprint {http://arxiv.org/abs/1812.06895} {arXiv:1812.06895
  [hep-th]} \BibitemShut {NoStop}%
\bibitem [{\citenamefont {Guevara}\ \emph {et~al.}(2021)\citenamefont
  {Guevara}, \citenamefont {Maybee}, \citenamefont {Ochirov}, \citenamefont
  {O'connell},\ and\ \citenamefont {Vines}}]{Guevara:2020xjx}%
  \BibitemOpen
  \bibfield  {author} {\bibinfo {author} {\bibfnamefont {A.}~\bibnamefont
  {Guevara}}, \bibinfo {author} {\bibfnamefont {B.}~\bibnamefont {Maybee}},
  \bibinfo {author} {\bibfnamefont {A.}~\bibnamefont {Ochirov}}, \bibinfo
  {author} {\bibfnamefont {D.}~\bibnamefont {O'connell}}, \ and\ \bibinfo
  {author} {\bibfnamefont {J.}~\bibnamefont {Vines}},\ }\href {\doibase
  10.1007/JHEP03(2021)201} {\bibfield  {journal} {\bibinfo  {journal} {JHEP}\
  }\textbf {\bibinfo {volume} {03}},\ \bibinfo {pages} {201} (\bibinfo {year}
  {2021})},\ \Eprint {http://arxiv.org/abs/2012.11570} {arXiv:2012.11570
  [hep-th]} \BibitemShut {NoStop}%
\bibitem [{\citenamefont {Guevara}(2021)}]{Guevara:2021yud}%
  \BibitemOpen
  \bibfield  {author} {\bibinfo {author} {\bibfnamefont {A.}~\bibnamefont
  {Guevara}},\ }\href@noop {} {\  (\bibinfo {year} {2021})},\ \Eprint
  {http://arxiv.org/abs/2112.05111} {arXiv:2112.05111 [hep-th]} \BibitemShut
  {NoStop}%
\bibitem [{\citenamefont {Chung}\ \emph {et~al.}(2019)\citenamefont {Chung},
  \citenamefont {Huang}, \citenamefont {Kim},\ and\ \citenamefont
  {Lee}}]{Chung:2018kqs}%
  \BibitemOpen
  \bibfield  {author} {\bibinfo {author} {\bibfnamefont {M.-Z.}\ \bibnamefont
  {Chung}}, \bibinfo {author} {\bibfnamefont {Y.-T.}\ \bibnamefont {Huang}},
  \bibinfo {author} {\bibfnamefont {J.-W.}\ \bibnamefont {Kim}}, \ and\
  \bibinfo {author} {\bibfnamefont {S.}~\bibnamefont {Lee}},\ }\href {\doibase
  10.1007/JHEP04(2019)156} {\bibfield  {journal} {\bibinfo  {journal} {JHEP}\
  }\textbf {\bibinfo {volume} {04}},\ \bibinfo {pages} {156} (\bibinfo {year}
  {2019})},\ \Eprint {http://arxiv.org/abs/1812.08752} {arXiv:1812.08752
  [hep-th]} \BibitemShut {NoStop}%
\bibitem [{\citenamefont {Maybee}\ \emph {et~al.}(2019)\citenamefont {Maybee},
  \citenamefont {O'Connell},\ and\ \citenamefont {Vines}}]{Maybee:2019jus}%
  \BibitemOpen
  \bibfield  {author} {\bibinfo {author} {\bibfnamefont {B.}~\bibnamefont
  {Maybee}}, \bibinfo {author} {\bibfnamefont {D.}~\bibnamefont {O'Connell}}, \
  and\ \bibinfo {author} {\bibfnamefont {J.}~\bibnamefont {Vines}},\ }\href
  {\doibase 10.1007/JHEP12(2019)156} {\bibfield  {journal} {\bibinfo  {journal}
  {JHEP}\ }\textbf {\bibinfo {volume} {12}},\ \bibinfo {pages} {156} (\bibinfo
  {year} {2019})},\ \Eprint {http://arxiv.org/abs/1906.09260} {arXiv:1906.09260
  [hep-th]} \BibitemShut {NoStop}%
\bibitem [{\citenamefont {Kim}\ and\ \citenamefont {Lee}(2023)}]{Kim:2023aff}%
  \BibitemOpen
  \bibfield  {author} {\bibinfo {author} {\bibfnamefont {J.-H.}\ \bibnamefont
  {Kim}}\ and\ \bibinfo {author} {\bibfnamefont {S.}~\bibnamefont {Lee}},\
  }\href@noop {} {\  (\bibinfo {year} {2023})},\ \Eprint
  {http://arxiv.org/abs/2301.06203} {arXiv:2301.06203 [hep-th]} \BibitemShut
  {NoStop}%
\bibitem [{\citenamefont {Kim}(2023)}]{Kim:2023vgb}%
  \BibitemOpen
  \bibfield  {author} {\bibinfo {author} {\bibfnamefont {J.-H.}\ \bibnamefont
  {Kim}},\ }\href@noop {} {\  (\bibinfo {year} {2023})},\ \Eprint
  {http://arxiv.org/abs/2309.11886} {arXiv:2309.11886 [hep-th]} \BibitemShut
  {NoStop}%
\bibitem [{\citenamefont {Arkani-Hamed}\ \emph {et~al.}(2020)\citenamefont
  {Arkani-Hamed}, \citenamefont {Huang},\ and\ \citenamefont
  {O'Connell}}]{Arkani-Hamed:2019ymq}%
  \BibitemOpen
  \bibfield  {author} {\bibinfo {author} {\bibfnamefont {N.}~\bibnamefont
  {Arkani-Hamed}}, \bibinfo {author} {\bibfnamefont {Y.-t.}\ \bibnamefont
  {Huang}}, \ and\ \bibinfo {author} {\bibfnamefont {D.}~\bibnamefont
  {O'Connell}},\ }\href {\doibase 10.1007/JHEP01(2020)046} {\bibfield
  {journal} {\bibinfo  {journal} {JHEP}\ }\textbf {\bibinfo {volume} {01}},\
  \bibinfo {pages} {046} (\bibinfo {year} {2020})},\ \Eprint
  {http://arxiv.org/abs/1906.10100} {arXiv:1906.10100 [hep-th]} \BibitemShut
  {NoStop}%
\bibitem [{\citenamefont {Aoude}\ and\ \citenamefont
  {Ochirov}(2021)}]{Aoude:2021oqj}%
  \BibitemOpen
  \bibfield  {author} {\bibinfo {author} {\bibfnamefont {R.}~\bibnamefont
  {Aoude}}\ and\ \bibinfo {author} {\bibfnamefont {A.}~\bibnamefont
  {Ochirov}},\ }\href {\doibase 10.1007/JHEP10(2021)008} {\bibfield  {journal}
  {\bibinfo  {journal} {JHEP}\ }\textbf {\bibinfo {volume} {10}},\ \bibinfo
  {pages} {008} (\bibinfo {year} {2021})},\ \Eprint
  {http://arxiv.org/abs/2108.01649} {arXiv:2108.01649 [hep-th]} \BibitemShut
  {NoStop}%
\bibitem [{\citenamefont {Penrose}\ and\ \citenamefont
  {Rindler}(2011)}]{Penrose:1985bww}%
  \BibitemOpen
  \bibfield  {author} {\bibinfo {author} {\bibfnamefont {R.}~\bibnamefont
  {Penrose}}\ and\ \bibinfo {author} {\bibfnamefont {W.}~\bibnamefont
  {Rindler}},\ }\href {\doibase 10.1017/CBO9780511564048} {\emph {\bibinfo
  {title} {{Spinors and Space-Time}}}},\ Cambridge Monographs on Mathematical
  Physics\ (\bibinfo  {publisher} {Cambridge Univ. Press},\ \bibinfo {address}
  {Cambridge, UK},\ \bibinfo {year} {2011})\BibitemShut {NoStop}%
\bibitem [{\citenamefont {Penrose}(1976)}]{penrosenlg}%
  \BibitemOpen
  \bibfield  {author} {\bibinfo {author} {\bibfnamefont {R.}~\bibnamefont
  {Penrose}},\ }\href {\doibase 10.1007/BF00763433} {\bibfield  {journal}
  {\bibinfo  {journal} {General Relativity and Gravitation}\ }\textbf {\bibinfo
  {volume} {7}},\ \bibinfo {pages} {171} (\bibinfo {year} {1976})}\BibitemShut
  {NoStop}%
\bibitem [{\citenamefont {Bardeen}\ and\ \citenamefont
  {Horowitz}(1999)}]{Bardeen:1999px}%
  \BibitemOpen
  \bibfield  {author} {\bibinfo {author} {\bibfnamefont {J.~M.}\ \bibnamefont
  {Bardeen}}\ and\ \bibinfo {author} {\bibfnamefont {G.~T.}\ \bibnamefont
  {Horowitz}},\ }\href {\doibase 10.1103/PhysRevD.60.104030} {\bibfield
  {journal} {\bibinfo  {journal} {Phys. Rev. D}\ }\textbf {\bibinfo {volume}
  {60}},\ \bibinfo {pages} {104030} (\bibinfo {year} {1999})},\ \Eprint
  {http://arxiv.org/abs/hep-th/9905099} {arXiv:hep-th/9905099} \BibitemShut
  {NoStop}%
\bibitem [{\citenamefont {Atanasov}\ \emph {et~al.}(2021)\citenamefont
  {Atanasov}, \citenamefont {Ball}, \citenamefont {Melton}, \citenamefont
  {Raclariu},\ and\ \citenamefont {Strominger}}]{Atanasov:2021oyu}%
  \BibitemOpen
  \bibfield  {author} {\bibinfo {author} {\bibfnamefont {A.}~\bibnamefont
  {Atanasov}}, \bibinfo {author} {\bibfnamefont {A.}~\bibnamefont {Ball}},
  \bibinfo {author} {\bibfnamefont {W.}~\bibnamefont {Melton}}, \bibinfo
  {author} {\bibfnamefont {A.-M.}\ \bibnamefont {Raclariu}}, \ and\ \bibinfo
  {author} {\bibfnamefont {A.}~\bibnamefont {Strominger}},\ }\href {\doibase
  10.1007/JHEP07(2021)083} {\bibfield  {journal} {\bibinfo  {journal} {JHEP}\
  }\textbf {\bibinfo {volume} {07}},\ \bibinfo {pages} {083} (\bibinfo {year}
  {2021})},\ \Eprint {http://arxiv.org/abs/2101.09591} {arXiv:2101.09591
  [hep-th]} \BibitemShut {NoStop}%
\bibitem [{Note3()}]{Note3}%
  \BibitemOpen
  \bibinfo {note} {This analysis is tied to position space so it is rather
  different than the approach of e.g. \cite {penrosenlg}.}\BibitemShut {Stop}%
\bibitem [{Note4()}]{Note4}%
  \BibitemOpen
  \bibinfo {note} {We warn that commutators can be interpreted as operations in
  twistor space but products such as $\gamma _{0}K_{i}$ cannot. However, this
  representation is useful to derive the algebra.}\BibitemShut {Stop}%
\bibitem [{\citenamefont {Mack}\ and\ \citenamefont
  {Todorov}(1969)}]{Mack:1969dg}%
  \BibitemOpen
  \bibfield  {author} {\bibinfo {author} {\bibfnamefont {G.}~\bibnamefont
  {Mack}}\ and\ \bibinfo {author} {\bibfnamefont {I.}~\bibnamefont {Todorov}},\
  }\href {\doibase 10.1063/1.1664804} {\bibfield  {journal} {\bibinfo
  {journal} {J. Math. Phys.}\ }\textbf {\bibinfo {volume} {10}},\ \bibinfo
  {pages} {2078} (\bibinfo {year} {1969})}\BibitemShut {NoStop}%
\bibitem [{\citenamefont {Griffiths}(1995)}]{griffiths:quantum}%
  \BibitemOpen
  \bibfield  {author} {\bibinfo {author} {\bibfnamefont {D.}~\bibnamefont
  {Griffiths}},\ }\href@noop {} {\emph {\bibinfo {title} {Introduction of
  Quantum Mechanics}}}\ (\bibinfo  {publisher} {Prentice Hall, Inc.},\ \bibinfo
  {year} {1995})\BibitemShut {NoStop}%
\bibitem [{\citenamefont {Adamo}\ \emph {et~al.}(2023)\citenamefont {Adamo},
  \citenamefont {Bogna}, \citenamefont {Mason},\ and\ \citenamefont
  {Sharma}}]{Adamo:2023fbj}%
  \BibitemOpen
  \bibfield  {author} {\bibinfo {author} {\bibfnamefont {T.}~\bibnamefont
  {Adamo}}, \bibinfo {author} {\bibfnamefont {G.}~\bibnamefont {Bogna}},
  \bibinfo {author} {\bibfnamefont {L.}~\bibnamefont {Mason}}, \ and\ \bibinfo
  {author} {\bibfnamefont {A.}~\bibnamefont {Sharma}},\ }\href@noop {} {\
  (\bibinfo {year} {2023})},\ \Eprint {http://arxiv.org/abs/2309.03834}
  {arXiv:2309.03834 [hep-th]} \BibitemShut {NoStop}%
\bibitem [{\citenamefont {Anninos}\ \emph {et~al.}(2009)\citenamefont
  {Anninos}, \citenamefont {Li}, \citenamefont {Padi}, \citenamefont {Song},\
  and\ \citenamefont {Strominger}}]{Anninos:2008fx}%
  \BibitemOpen
  \bibfield  {author} {\bibinfo {author} {\bibfnamefont {D.}~\bibnamefont
  {Anninos}}, \bibinfo {author} {\bibfnamefont {W.}~\bibnamefont {Li}},
  \bibinfo {author} {\bibfnamefont {M.}~\bibnamefont {Padi}}, \bibinfo {author}
  {\bibfnamefont {W.}~\bibnamefont {Song}}, \ and\ \bibinfo {author}
  {\bibfnamefont {A.}~\bibnamefont {Strominger}},\ }\href {\doibase
  10.1088/1126-6708/2009/03/130} {\bibfield  {journal} {\bibinfo  {journal}
  {JHEP}\ }\textbf {\bibinfo {volume} {03}},\ \bibinfo {pages} {130} (\bibinfo
  {year} {2009})},\ \Eprint {http://arxiv.org/abs/0807.3040} {arXiv:0807.3040
  [hep-th]} \BibitemShut {NoStop}%
\bibitem [{\citenamefont {Zimmerman}\ and\ \citenamefont
  {Shahir}(1989)}]{Zimmerman:1989kv}%
  \BibitemOpen
  \bibfield  {author} {\bibinfo {author} {\bibfnamefont {R.~L.}\ \bibnamefont
  {Zimmerman}}\ and\ \bibinfo {author} {\bibfnamefont {B.~Y.}\ \bibnamefont
  {Shahir}},\ }\href {\doibase 10.1007/BF00758986} {\bibfield  {journal}
  {\bibinfo  {journal} {Gen. Rel. Grav.}\ }\textbf {\bibinfo {volume} {21}},\
  \bibinfo {pages} {821} (\bibinfo {year} {1989})}\BibitemShut {NoStop}%
\bibitem [{\citenamefont {Wu}\ and\ \citenamefont {Yang}(1976)}]{Wu:1976ge}%
  \BibitemOpen
  \bibfield  {author} {\bibinfo {author} {\bibfnamefont {T.~T.}\ \bibnamefont
  {Wu}}\ and\ \bibinfo {author} {\bibfnamefont {C.~N.}\ \bibnamefont {Yang}},\
  }\href {\doibase 10.1016/0550-3213(76)90143-7} {\bibfield  {journal}
  {\bibinfo  {journal} {Nucl. Phys. B}\ }\textbf {\bibinfo {volume} {107}},\
  \bibinfo {pages} {365} (\bibinfo {year} {1976})}\BibitemShut {NoStop}%
\bibitem [{\citenamefont {Shnir}(2005)}]{Shnir:2005vvi}%
  \BibitemOpen
  \bibfield  {author} {\bibinfo {author} {\bibfnamefont {Y.~M.}\ \bibnamefont
  {Shnir}},\ }\href {\doibase 10.1007/3-540-29082-6} {\emph {\bibinfo {title}
  {{Magnetic Monopoles}}}},\ Text and Monographs in Physics\ (\bibinfo
  {publisher} {Springer},\ \bibinfo {address} {Berlin/Heidelberg},\ \bibinfo
  {year} {2005})\BibitemShut {NoStop}%
\bibitem [{\citenamefont {Bethe}\ and\ \citenamefont
  {Salpeter}(1957)}]{Bethe:1957ncq}%
  \BibitemOpen
  \bibfield  {author} {\bibinfo {author} {\bibfnamefont {H.~A.}\ \bibnamefont
  {Bethe}}\ and\ \bibinfo {author} {\bibfnamefont {E.~E.}\ \bibnamefont
  {Salpeter}},\ }\href {\doibase 10.1007/978-3-662-12869-5} {\emph {\bibinfo
  {title} {{Quantum Mechanics of One- and Two-Electron Atoms}}}}\ (\bibinfo
  {year} {1957})\BibitemShut {NoStop}%
\end{thebibliography}%
	
\end{document}